# Multicast-based Architecture for IP Mobility: Simulation Analysis and Comparison with Basic Mobile IP

*Ahmed Helmy*
*Electrical Engineering Department, University of Southern California*
*helmy@ceng.usc.edu*

**Abstract**

*With the introduction of a newer generation of wireless devices and technologies, the need for an efficient architecture for IP mobility is becoming more apparent. Several architectures have been proposed to support IP mobility. Most studies, however, show that current architectures, in general, fall short from satisfying the performance requirements for wireless applications, mainly audio. Other studies have shown performance improvement by using multicast to reduce latency and packet loss during handoff. In this study, we propose a multicast-based architecture to support IP mobility. We evaluate our approach through simulation, and we compare it to mainstream approaches for IP mobility, mainly, the Mobile IP protocol. Comparison is performed according to the required performance criteria, such as smooth handoff and efficient routing.*

*Our simulation results show significant improvement for the proposed architecture. On average, basic Mobile IP consumes almost twice as much network bandwidth, and experiences more than twice as much end-to-end and handoff delays, as does our proposed architecture.*

*Furthermore, we propose an extension to Mobile IP to support our architecture with minimal modification.*

## 1. Introduction

The recent advances in wireless communication technology and the unprecedented growth of the Internet have paved the way for wireless networking and IP mobility. Unlike conventional wired networks, wireless networks possess different channel characteristics and mobility dynamics that render network design and analysis more challenging. Performance during *handoff* - where the mobile moves from one cell, or coverage area, to another - is a significant factor in evaluating wireless networks. In addition, route efficiency is a measure to evaluate the impact of the mobility architecture on the network, in terms of resources consumed, or overhead incurred.

IP-multicast provides efficient algorithms for multiple packet delivery. It also provides location-independent group addressing. The receiver-initiated approach for IP-multicast enables new receivers to join to a nearby branch of an already established multicast tree. Hence, IP-multicast provides a scalable infrastructure for efficient, location-independent, packet delivery.

In our study, we attempt to leverage off of the merits of IP-multicast to lay out our architecture. Earlier studies[9][11] have suggested that using multicast principles may improve performance of IP mobility. However, these studies did not quantify this improvement in the wide-area network or Internet contexts. We design and present a multicast-based architecture according to clear design requirements, and evaluate our architecture with respect to these requirements using large-scale simulations. We focus on performance during handoff and route efficiency as the primary metrics for evaluation.

The intuition behind our proposal is simple. As the mobile node roams across the network, we want packets destined to it to *follow it* throughout its movement. Imagine a dynamic distribution tree with branches reaching all locations visited by the mobile during its journey. These branches constitute the shortest paths from the packet source to each of the visited locations. The tree is dynamic such that the branches grow and shrink to reach the mobile node as necessary, when necessary. This architecture is realized by having the correspondent node send packets to a multicast address, to which the mobile node joins from each location visited. In this paper, we describe the mechanisms involved in realizing such architecture. In addition, we analyze the performance of these mechanisms through simulation and compare it to Mobile IP[4].

Our simulation results show that, on average, basic Mobile IP consumes almost twice as much network bandwidth, and experiences more than twice as much end-to-end and handoff delays, as does our proposed architecture. For systems already using Mobile IP with route optimization, we propose a minimal extension to support the architecture presented in this paper. The multicast address assigned to the mobile node can be sent to the correspondent node through a *binding update* in the start-up phase of communication.

The rest of this paper is organized as follows. Section 2 outlines the design requirements. Architectural overview is presented in Section 3. In Section 4 we present our performance evaluation, while Section 5 discusses design issues and alternatives. Section 6 presents related work. We conclude in Section 7 and present directions for future work.

## 2. Design Requirements

The main requirement of this architecture is to provide an efficient mechanism for IP mobility. Efficiency is measured in terms of performance, compatibility and applicability.

### A. *Performance requirements*

The proposed architecture should satisfy the following performance requirements.

I. Smooth handoff:

A general requirement for IP mobility is to provide 'smooth' handoff. Handoff *smoothness* can be measured by several criteria, such as delay, jitter, data loss and communication overhead during handoff. The efficiency of handoff depends heavily on the specific application. For example, audio in general is tolerant to loss, but has stringent delay requirements. Handoff delay is a function of the number of hops added to the data path during handoff[1]. We strive reduce handoff latency for much of the operating conditions studied.

II. Efficient routing:

One major drawback of the Mobile IP protocol is triangle routing, where packets from the correspondent node travel to the home agent (in the mobile's home network) before being tunneled to the mobile node. This approach: (a) consumes a lot of network resources, (b) is more susceptible to network partition[2] and (c) degrades the performance perceived by the end applications. It further complicates the handoff process. In our proposed mechanisms we attempt to avoid the above drawbacks. Routing efficiency may be measured in terms of number of hops traversed by data packets from the correspondent node (CN) to the mobile node (MN)[3], as well as the overall network bandwidth consumed due to the mobility architecture.

III. Low waste of network and RF bandwidth:

Bandwidth of wireless links (such as radio frequency 'RF' links) is a scarce resource. Some approaches attempt to minimize delay and loss during handoff by multiple packet forwarding. Hence, these approaches are likely to waste more bandwidth than others are. We design protocol mechanisms that conserve network bandwidth in general, and reduce wasted RF bandwidth during handoff.

### B. *Applicability and Compatibility requirements*

The basic version of our architecture assumes multicast capability in the Internet[4]. In our simulations we mainly use the Protocol Independent Multicast-Sparse Mode (PIM-SM)[3] as the multicast routing protocol. Our architecture also assumes that each mobile node is assigned a multicast identifier. This identifier is typically a multicast address. For IPv6, we do not expect this to be a problem[5].

## 3. Architectural Overview

In order to provide mobility, the packets sent to the mobile node (MN) need to be forwarded to every location visited by the MN. Forwarding takes place according to the temporal and spatial pattern of movement. One may view this problem as follows. The set of locations that the mobile will visit may be viewed as a *group* of receivers, to which the packets should be delivered. The temporal *pattern* by which packets are delivered to these receivers represents the temporal component of the movement.

---

[1] For our architecture, this will be the number of links added from the new location to establish a branch from the existing multicast delivery tree.
[2] We call this the "3rd party dependency" problem.
[3] We define the ratio *r* as the number of hops traversed by the data packets from the CN to the MN in the basic Mobile IP architecture to those traversed in our architecture.
[4] Although such infrastructure is not as yet fully deployed, it is envisioned that in the few years to come multicast will be ubiquitous.
[5] For IPv4, however, this may prove to be problematic due to the limited multicast address space (class D IP-addresses). This is still an open issue under investigation.

Instead of sending their packets to a unicast address, nodes wishing to send to the MN send their packets to a *multicast group* address. The MN, throughout its movement, would join this multicast group through the locations it visits[6]. Because the movement will be to a geographical vicinity, it is highly likely that the join from the new location (to which the mobile has recently moved) will traverse a small number of hops to reach the multicast distribution tree (already established to the previous location of the mobile node). Hence, performance during handoff (in terms of latency and packet loss) will be improved drastically. In this section, we discuss how the main components of our architecture interact to realize the desired behavior.

## A. Dynamics of packet distribution

Optimally, the packets destined to the mobile node should be delivered over the shortest path, traversing the minimum number of links and experiencing minimum delay. The packets should traverse only those links that lead to the current location of the mobile node, but also should be delivered to the new location with minimum handoff latency. This may be achieved by redirecting the packet delivery tree (of the old location) to grow in the direction of minimum distance to the new location. That is, a branch should be established from the new location of the mobile node to the nearest point of the delivery tree. This is illustrated in Figure 1. As the mobile node continues to move, branches are established to deliver packets to the new location, while other branches, those that no longer lead to the mobile node, are torn down and pruned.

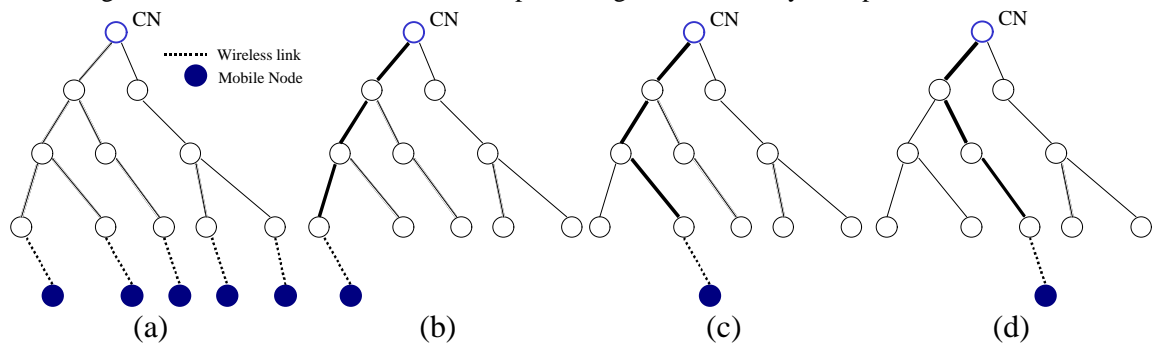

**Figure 1.** Architectural concept: (a) all the locations visited by the mobile are considered part of the distribution tree (at some point), (b) when a mobile moves to a certain location, only that location becomes part of the tree (as shown by bold lines), when the mobile moves to a new location, as in (c) and (d) the distribution tree changes to deliver packets to the new location.

## B. Establishing the delivery tree

Initially, the MN is assigned a multicast address *'G'*. The CN sends its packets to *G*. To establish the delivery tree from the CN to the MN, the MN sends a source-specific *(CN,G)* join message towards the CN[7], as in Figure 2 (a). As the MN moves and connects to another location[8], it joins *(CN,G)* through the new location, as in Figure 2 (b). The join is forwarded upstream (according to the multicast routing protocol) towards CN. When the join reaches the nearest point of the multicast tree, the join process is complete and the packets start flowing down the newly established branch towards the MN. Upon receiving data packets from the new location, the MN issues a prune message to the old location. The prune message tears down the branches that no longer lead to the MN[9].

---

[6] Dynamics of joining and leaving the group simulate the movement of the mobile node.
[7] At this phase of the protocol, we assume that a start-up phase has already been completed, through which the MN is assigned a multicast address and gains knowledge of the CN, and the CN gains knowledge of the MN's multicast address. We discuss this start-up phase later on in this paper. We also assume that the multicast protocol allows explicit joining and pruning and supports source-specific trees, similar to PIM-SM[3] or BGMP[26] capabilities, and that the host interface allows for source-group specific joins, as in IGMPv3[25].
[8] By location, we mean a new point of attachment that leads to change in the first hop router. For simplicity, without loss of generality, we assume that each base station is a router.

## C. Smooth Handoff

In general, at any point in time, the MN should accept packets from only one location. However, during transient movement, the MN may be joined to *G* through multiple locations. The dynamics of joining and leaving/pruning *G* during handoff directly affect handoff latency and smoothness. To allow a smooth handoff, the MN should not prune the old location until/unless it starts getting packets from the new location. This is illustrated in Figure 3. To further ensure smoothness and to conserve RF bandwidth, mechanisms should be designed to reduce join latency and leave latency. We discuss this issue in Section 5.

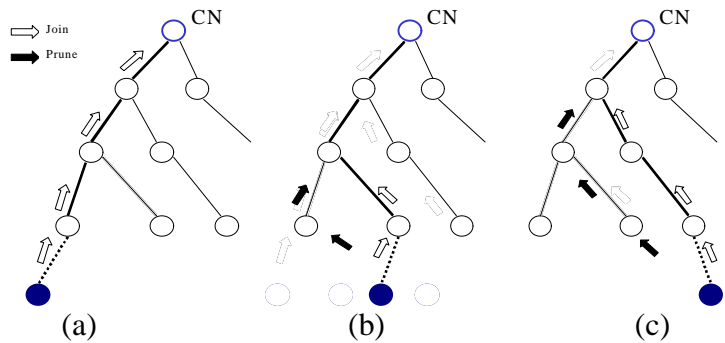

**Figure 2.** The distribution tree adapts to movement of the MN: (a) MN joins towards CN. As the MN moves, as in (b) and (c), the MN joins the distribution tree through the new location and prunes through the old location.

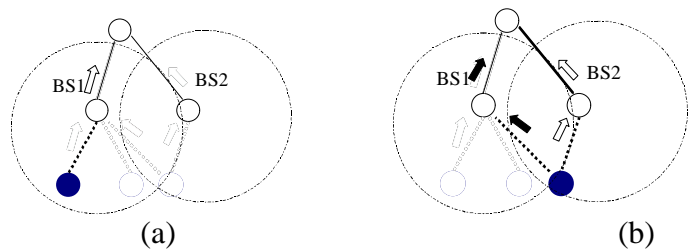

**Figure 3.** Smooth handoff between neighboring cells: (a) MN joins through BS1, (b) as the MN moves into BS2's covering region it joins through BS2, when it starts getting packets from BS2 it sends a leave/prune through BS1.

## 4. Performance Evaluation

Most studies conducted on IP mobility use very limited topologies and scenarios to evaluate their architecture, focusing only on handoff behavior over LANs[9][10][11][12][13][14]. We would like to examine the behavior of such architectures (basic Mobile IP[4], in specific), in addition to the one proposed here, in the context of wide area networks. We use simulation to perform the comparison. The simulation scenarios consist mainly of the topology and movement models.

I.   *Topology:* The topology model mainly defines the number of nodes and their link connectivity. Several methods may be used to generate simulation topologies. We have chosen three methods that we believe result in a good variety of topology samples. These three methods include two topology generators (GT-ITM[16] and Tiers[23]) and a set of real measured topologies. Table 1 lists the topologies used in our simulations[10]. The topologies include 47 to 5000 nodes with different degrees of connectivity. A couple of these topologies are shown in Figure 4 and 5[11].

---

[9] If upon receiving the first few packets from the new location the MN cannot communicate with the old location, then the prune should be triggered through another implicit mechanism, such as a time-out mechanism, or through a triggered message from the old location upon detection of loss of connectivity to the MN.

[10] We included topologies used in previous studies on the Internet[18][19]. ARPA is the original ARPANET topology, the Mbone topologies are based on measurements by the SCAN project at USC/ISI (www.isi.edu/scan), and the AS

II.    *Movement:* The movement pattern defines the sequence of nodes visited by the mobile node (MN) throughout the simulation. We have used three movement patterns as follows. The first allows the MN to visit any node in the topology randomly, we call this movement pattern *random*. The second allows the MN to visit only directly-connected nodes in the next movement step. The next neighbor is chosen randomly from the set of directly connected nodes. We call the second movement pattern *neighbor*. The third movement pattern allows the MN to connect randomly to only one of 6 nodes that are likely to fall within the same cluster as the MN[12]. This movement pattern is called *cluster*. Examples of the cluster movement are shown in Figure 6 for 4 different simulation runs, each with 100 movements.

| name | nodes | links | avg. deg | name | nodes | links | avg. deg | name | nodes | links | avg. deg |
|---|---|---|---|---|---|---|---|---|---|---|---|
| r50 | 50 | 217 | 8.68 | ts150 | 150 | 276 | 3.71 | ts1008_3 | 1008 | 3787 | 7.51 |
| r100 | 100 | 950 | 19 | ts200 | 200 | 372 | 3.72 | ti1000 | 1000 | 1405 | 2.81 |
| r150 | 150 | 2186 | 29.15 | ts250 | 250 | 463 | 3.72 | ti5000 | 5000 | 7084 | 2.83 |
| r200 | 200 | 3993 | 39.93 | ts300 | 300 | 559 | 3.73 | Mbone_1 | 3927 | 7555 | 3.85 |
| r250 | 250 | 6210 | 49.68 | ts1000 | 1000 | 1819 | 3.64 | Mbone_2 | 4179 | 8549 | 4.09 |
| ts50 | 50 | 89 | 3.63 | ts1008_1 | 1008 | 1399 | 2.78 | AS | 4830 | 9077 | 3.76 |
| ts100 | 100 | 185 | 3.7 | ts1008_2 | 1008 | 2581 | 5.12 | ARPA | 47 | 68 | 2.89 |

**Table 1.** Topologies used in the simulation (r: flat random, ts: transit-stub, ti: Tiers)

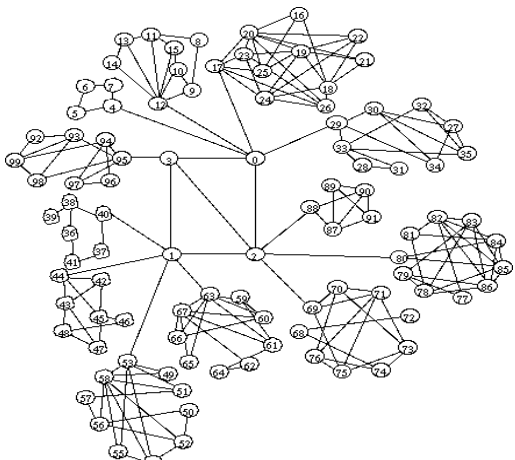

**Figure 4.** 100 node transit-stub topology (ts100)[13]

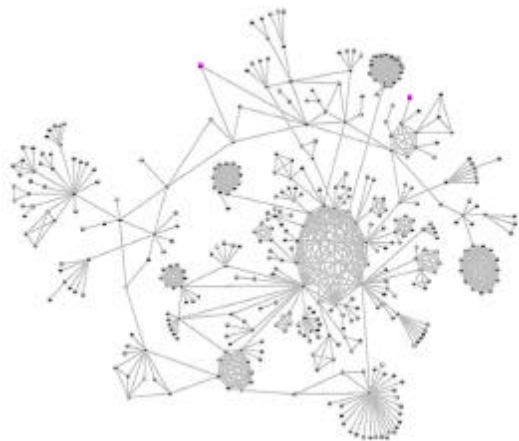

**Figure 5.** Map of the MBone[14]

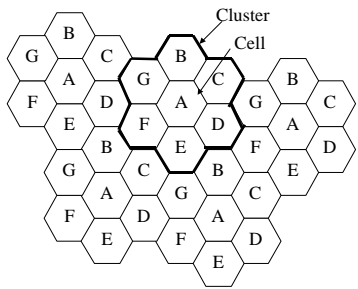

(a)

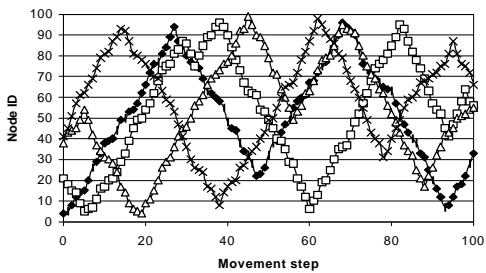

(b)

**Figure 6.** (a) A cluster of cells (in bold) is replicated over the coverage area. Cluster size is 7. When moving, MN in cell 'A' may visit one of 6 neighboring cells within the same cluster. (b) Cluster movement Patterns for ts100.

---

topology is provided by NLANR (moat.nlanr.net/Routing). For the GT-ITM generated topologies, we used flat random and transit-stub (or clustered) topologies.
[11] Note the obvious clustering in these topologies.
[12] This is especially meaningful in the transit-stub and clustered topologies, where nodes are numbered sequentially in a cluster, so node numbering has geographical significance. We generally assume that this is a cellular system with 7 cell reuse architecture, where each node in the topology is a base station for a cell, hence the 6 node figure. See Figure 6.
[13] This topology was used in a previous study[24].
[14] This is a visualization of an actual run of the SCAN project at USC/ISI (www.isi.edu/scan/scan.html).

### 4.1. Simulation

We use the VINT[20] tool kit for simulation, including the network simulator (ns)[21] and the network animator (nam)[22]. The unicast routing we use is Dijkstra's algorithm, and the multicast routing protocol is the centralized PIM-SM[3] with the SPT switch to enable source-based trees[15].

For each simulation run, a correspondent node (CN) and a home agent (HA) are chosen randomly, and 100 nodes are chosen as movement steps. The 100 nodes are chosen according to one of three movement patterns defined above. These are the nodes visited by the mobile node (MN) during the movement. For each movement pattern, simulation is repeated 10 times with different random number generator seeds.

We measure the number of hops traversed by data packets in our approach and in the basic Mobile IP approach. In addition, we measure the number of hops added to the multicast tree during movement from one node to another (i.e., during handoff).

### 4.2. Simulation Results

We measure the total number of links traversed during the simulation, and the ratio between the paths[16] taken by data packets in MIP and those taken in our architecture. We call this ratio '$r$'. Figure 7 (a) shows $r = (A+B)/C$, where $A$ is the unicast path (i.e., number of hops) from CN to HA, $B$ is the unicast path from HA to MN, and $C$ is the multicast path from CN to MN. We also measure, for our architecture, the number of hops added to the multicast distribution tree with every move. This number is denoted as '$L$', and is shown in Figure 7 (b). To illustrate, example measurement for the ts100 topology is given in Figure 8. As shown, a move from node 66 to node 72 produced 7 added links. By examining the ts100 topology (Figure 4) we see that such a move occurs between clusters[17]. We also measure the ratio '$B/L$' as a measure of the handoff latency between Mobile IP and our architecture.

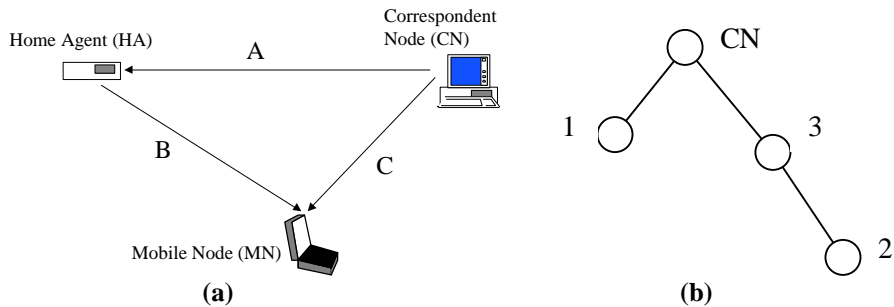

**Figure 7.** (a) Ratio '$r$'= $(A+B)/C$. (b) As the MN moves from node 1 to 2, the number of added links '$L$' is 2. As it moves from 2 to 3 there are no added links ($L=0$).

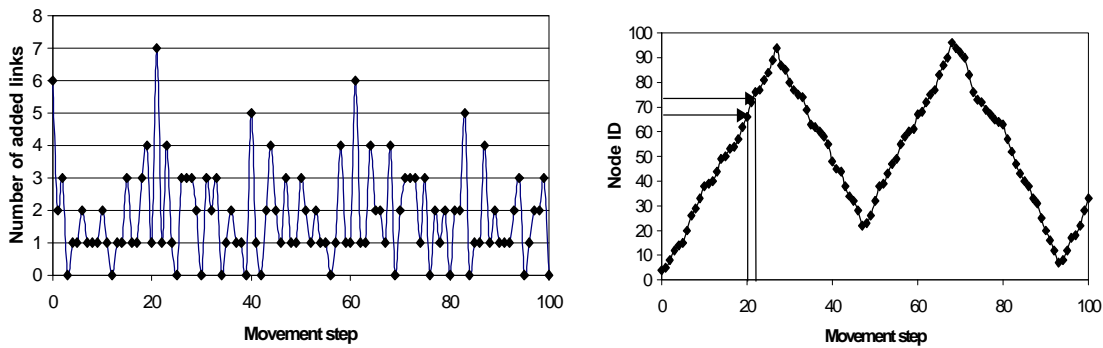

**Figure 8.** Number of added links. Arrows point to the move with maximum added links.

---

[15] In effect, the RPF source-based tree built by PIM-SM is the same as the unicast shortest path under assumption of path symmetry.
[16] The path length is measured in terms of number of hops.
[17] Other peaks include moves from node 55 to 48, 67 to 68, and 52 to 47 (all are inter-cluster movements).

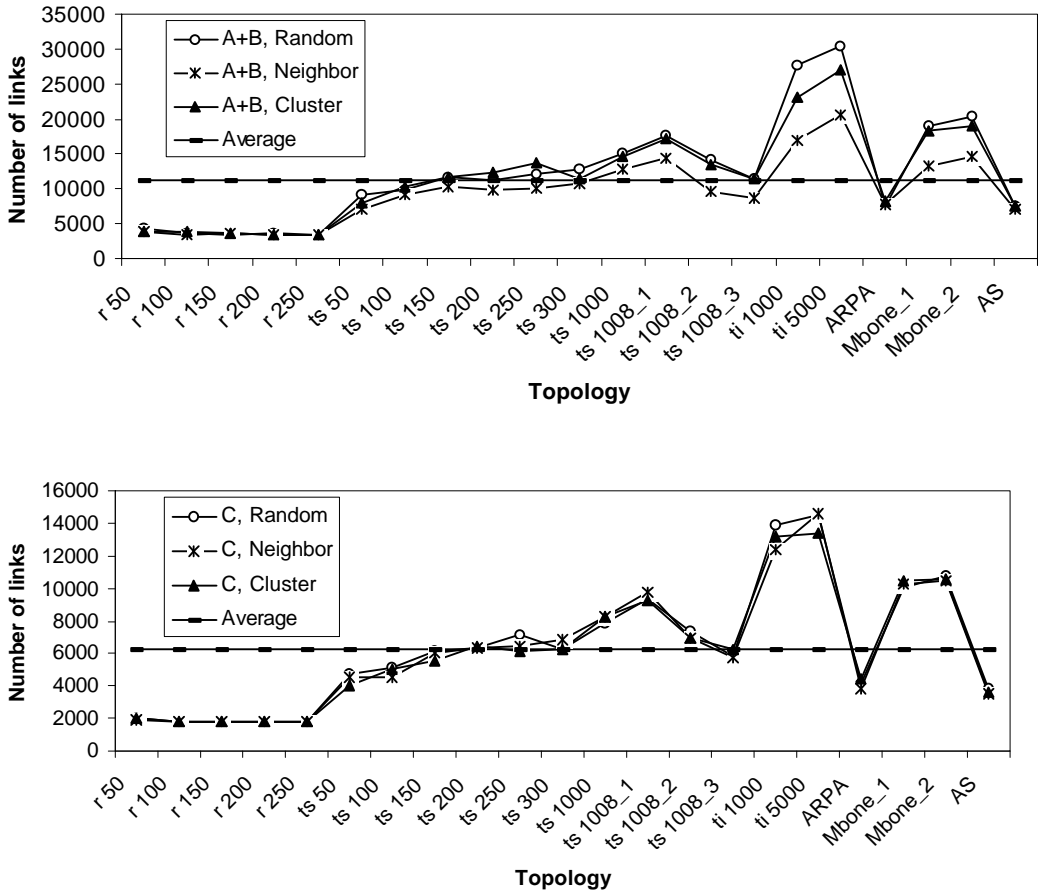

**Figure 9.** Total links traversed in Mobile IP 'A+B', and the total links traversed in our architecture 'C'.

Statistics are computed for 1000 samples (100 movements and 10 runs) for each topology, then averaged across same type topologies (e.g., random, transit-stub). Our results indicate that for Mobile IP, data may traverse up to '29'[18] times the number of hops as in our architecture with an average of '2.11' times more. Our results also show that for the multicast-based mobility architecture the average number of hops added to the multicast tree[19] during handoff was '2.51'. We discuss the simulation results in detail in the rest of this section.

### 4.3. Network links traversed

The total number of links traversed by the data packets is one metric to measure the consumed network resources. Here, we present the total number of links traversed throughout our simulations. For our proposed architecture the total links are denoted by 'C', while the total links traversed by Mobile IP are denoted by 'A+B', both shown in Figure 9. In general, the total links traversed is quite similar for all types of movement models[20]. The average number of links traversed across all topologies in our architecture is 6,208 links (for 1,000 samples), while the average for Mobile IP is 11,157. That is, on average, Mobile IP consumes almost twice as much network resources as does our architecture, for the same scenarios.

---

[18] This occurred for the ti1000 topology, during the *random* movement pattern.
[19] Our results show that the cumulative number of added links is an upper bound on the cumulative number of removed links. *You cannot remove more links than those already added.*
[20] Although it is a bit lower for the *neighbor* movement in case of Mobile IP, because it is the most constrained among the three movement models used.

## 4.4. Ratio 'r'

For different topologies and movement patterns the ratio *'r'* is measured as the metric for routing efficiency. This ratio mainly compares the path lengths from the CN to the MN[21]. The measurement results of the ratio *'r'* are given in Figure 10. The average ratio across topology types ranged from 1.56 to 2.42[22], while the 90th percentile point ranged from 2.13 to 5[23]. The maximum ratio ranges from 4 to 26[24]. The overall average of *r* is '2.11'. Hence, on average, the end-to-end delay experienced by Mobile IP is more than twice as much the delay experienced by our proposed architecture.

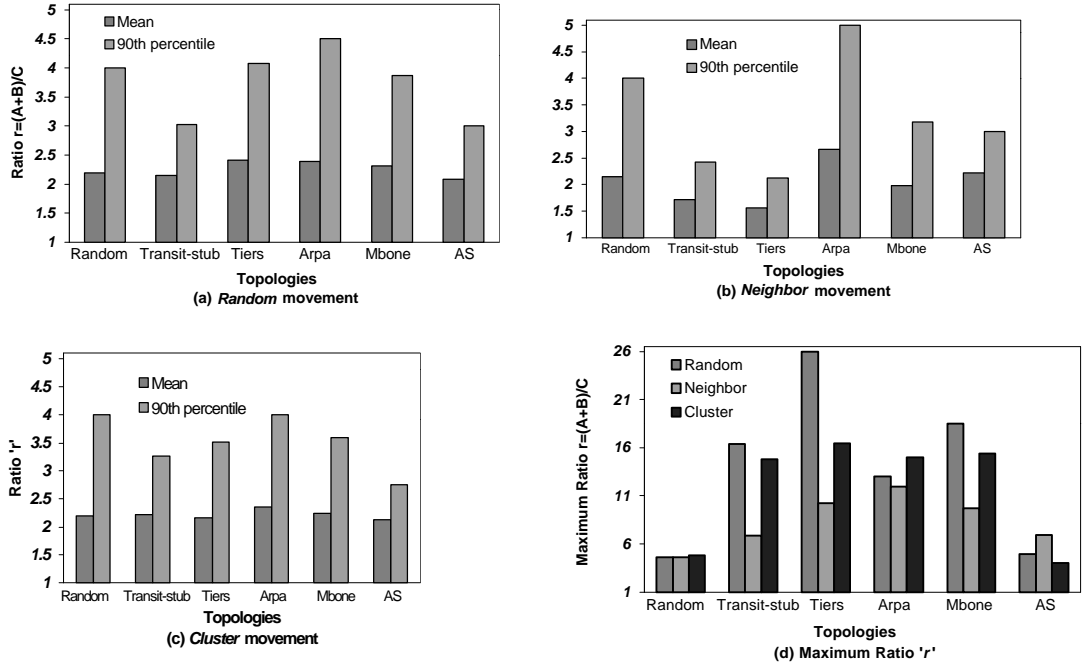

**Figure 10.** Ratio *'r'* for the different topologies and movement patterns.

In terms of sensitivity to topology size, we found that the ratio *'r'* is relatively insensitive to topology size. This is clearly shown in Figure 11, as average *r* is almost constant for the different topology sizes.

## 4.5. Added links 'L'

In our proposed architecture, handoff delay is directly related to *L*, since *L* represents the number of links traversed by the join message to pull the data packets down to the new location. Simulation estimates for *L* are shown in Figure 12. The movement experiencing the least *L* is *neighbor* (with overall average of 1.18), then *cluster* (with average of 2.38), and then *random* (with average of 3.97). It is clear that the more restrictive the movement model, the lower the value of *L*. The overall average number of added links *L* is 2.51 links. For most practical purposes, we believe that *cluster* and *neighbor* models are more suitable -than *random*- for representing mobile movement. For *cluster* and *neighbor* movements, the average is 1.78 links. Also, of practical relevance are the Mbone topologies simulated with *cluster* and *neighbor* movements. These scenarios have an average of about 2.5 links.

---

[21] This relates directly to the end-to-end delays experienced by the data packets, as opposed to the network bandwidth consumed that is represented by the total links traversed.

[22] Both the minimum and the maximum averages for *r* were recorded for the Tiers topologies for the *random* and *neighbor* movements, respectively.

[23] The maximum 90th percentile point was recorded for the ARPA topology for the *neighbor* movement, while the minimum was recorded for the Tiers topology for the *neighbor* movement.

[24] The least maximum ratio of 4 was recorded for the AS topology with cluster movement, where the maximum ratio of 26 was recorded for the Tiers topologies with random movement. [The maximum recorded was 29 for the ti1000, but here we report the average of each topology type.]

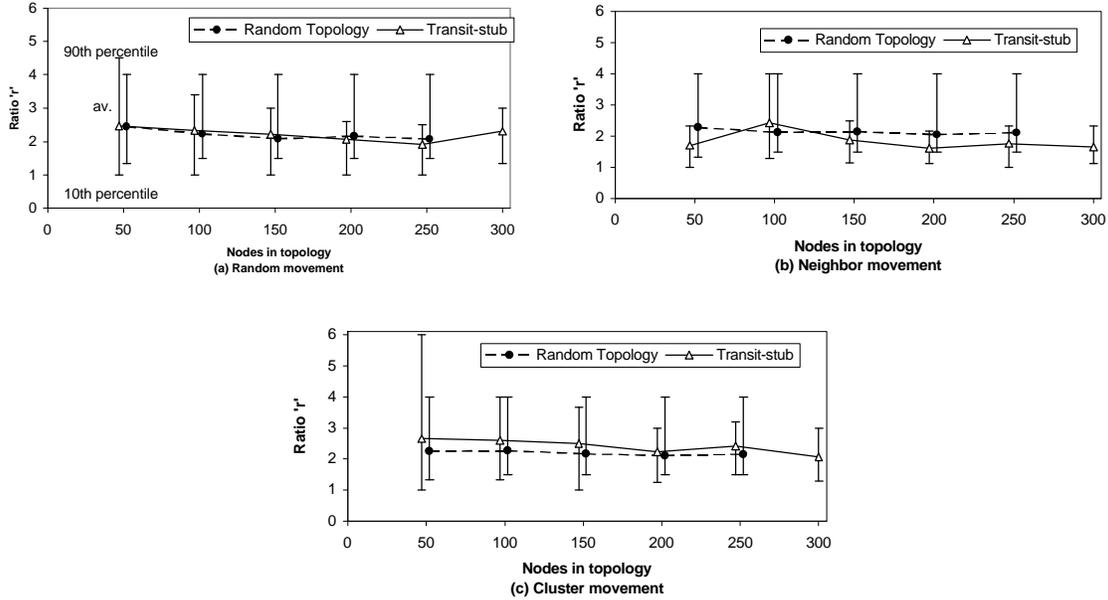

**Figure 11** Ratio *'r'* for various random and transit-stub topology sizes. *r*, is relatively the same for the different topology sizes (given in number of nodes).

In terms of sensitivity of *L* to topology size, we find that *L* is relatively insensitive to topology size (as shown in Figure 13), with sensitivity decreasing, as the movement model becomes more constraint.

### 4.6. *B/L* Ratio

As noted above, in our proposed architecture, handoff delay is a function of round trip time (*rtt*) to establish the new branch of the tree, which is, in turn, a function of *L*[25]. For Mobile IP, handoff delay is a function of the delay to register the new care-of-address and have the packets flow to the new location from the HA (i.e., handoff delay is function of *rtt* from the MN to the HA). In turn, this is a function of *B*, as shown in Figure 7. We define the ratio '*B/L*'[26] as one measure to compare handoff latency between our proposed architecture and Mobile IP. *B/L* ratio measures for our simulations are given in Figure 14. The ratio ranges from 1.04 (for random topologies with cluster movement) to 4.32 (for tiers topologies with neighbor movement), with an average of '2.31' for all topologies and movements[27]. Hence, on average, the handoff latency due to the Mobile IP architecture is more than twice as much that for our proposed architecture.

## 5. Design Issues and Alternatives

Thus far, we have outlined our proposed architecture and presented extensive simulation results for its performance in the wide-area context. However, several design issues need to be discussed to make our architecture more complete. In this section, we discuss some of these issues briefly.

### 5.1. Start-up Phase

In order to send packets to the multicast address of the mobile node, the correspondent node needs to obtain this multicast address. One may argue that such an address may be obtained in a way similar to obtaining the unicast address of a mobile node in the conventional Mobile IP (e.g., through DNS lookup, or otherwise). However, we do not assume such a service to be available, but provide an alternative design, one that requires minor modification to MIP with route optimization[5][6]. We propose that the multicast address be conveyed to the CN by the MN using *binding update*. However, unlike MIP with route optimization, the binding update occurs only once during the initial establishment of communication, not with every move[28].

---

[25] Handoff delay is also a function of the delay experienced over the wireless link and the detailed protocol mechanisms that are used, among other things. The detailed simulation is subject to future work. In this study, however, we focus on delays due to the general architecture (not mechanistic details), of which we believe *L* and *B* to be major components.

[26] Since *L* may become zero in some movement steps, we calculate average *L* and average *B* for all movement steps within a simulation, then we get '*av. B/ av. L*' ratio.

[27] This ratio '*B/L*' goes up to 2.7 if we exclude the flat random topologies, which are the least practical of the topologies used in our study.

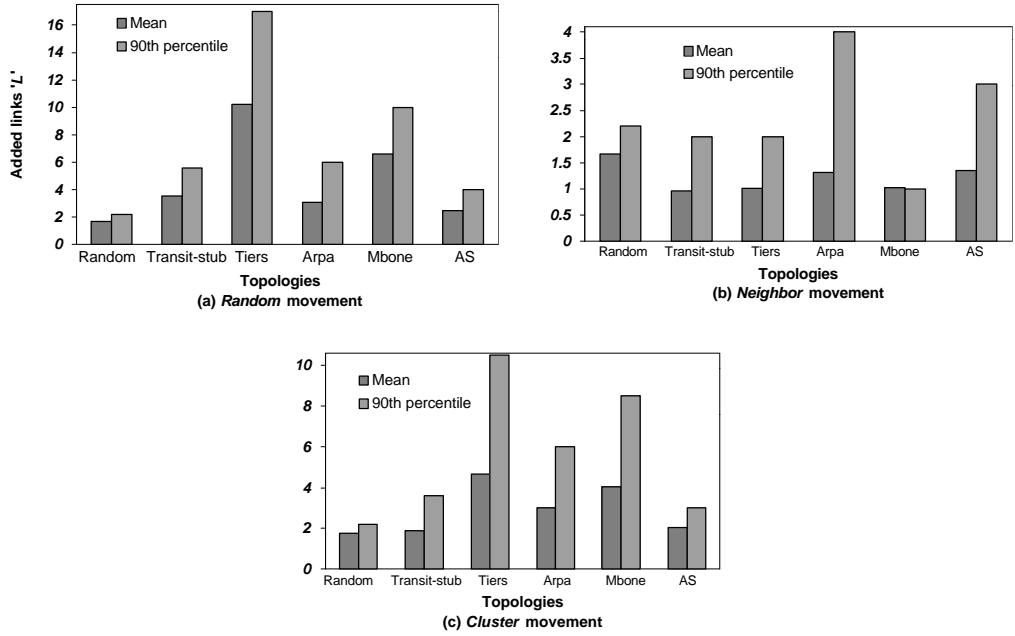

**Figure 12.** Added links '*L*' (average and 90[th] percentile) for different topologies and movement patterns.

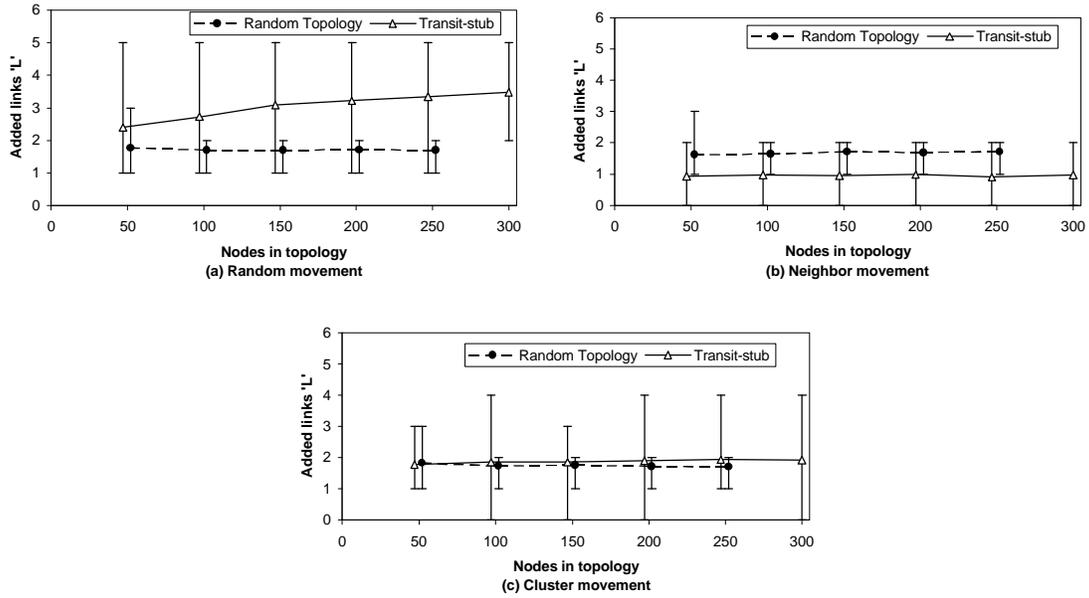

**Figure 13.** Added links '*L*' for various random and transit-stub topology sizes.

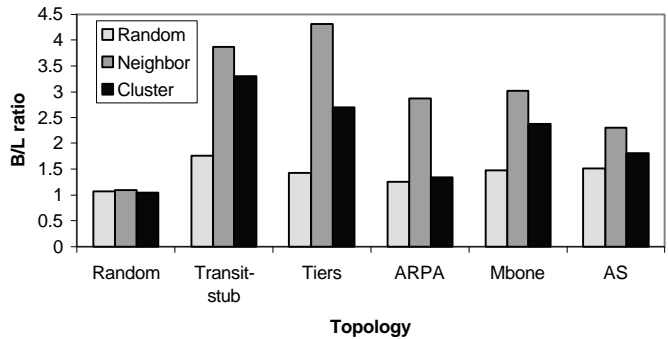

**Figure 14** Average *B/L* ratio for the different simulated topologies and movement models

## 5.2. Membership Host Interface

As the MN moves to a new location, it needs to join the multicast group *G* to which it is assigned, and it needs to specify the CN(s) as the source(s) to which it wishes to join. Hence, the membership host interface needs to support (*S,G)* joins, that are directed towards the source directly. Work on IGMPv3[25] may provide this support. However, IGMP was not designed specifically for wireless links. In addition, in our architecture, the MN is the only member of group *G*, while IGMP was designed to provide for suppression in case of multiple receivers on the LAN. We propose to design another protocol, that is more suitable for wireless links than IGMP, to provide the required functionality.

## 5.3. Reducing Join and Leave Latency

Several multicast routing protocols[3] use soft state mechanisms to achieve robustness. Soft state mechanisms use periodic refresh timers, and hence if a join is lost it will be recovered after the refresh period[29]. To maintain robustness, we propose to still use the soft state approach. However, we also propose to use another mechanism when the state is first created or removed to reduce the join and leave latency, respectively. Either an ack'ed mechanism, or, if even less latency is required, then we propose to send three back-to-back join (prune) messages upon creation (removal) of the state to the upstream routers.

## 5.4. Caching vs. Advanced Joins

In order to reduce number of packets lost during handoff, some schemes suggest to use caching. In MIP with route optimization, it is proposed to cache the packets (that otherwise would be lost during handoff) in the previous foreign agent, and then forward these packets to the new foreign agent after handoff is complete. Instead, we propose to use advanced joining, where the MN chooses the next point of attachment (based on signal strength measurements, for example), and sends an 'advance' join to that point. Advance joins are sent such that packets are already forwarded to the new location by the time the MN has relocated. We argue that forwarding cached information introduces delay, jitter, and out-of-order packets, and hence is unsuitable for delay-sensitive applications. For applications requiring reliability, then re-transmission should be done by higher layer protocols.

## 5.5. Scalability and Robustness

In this study, we have analyzed several performance aspects of our proposed architecture in a wide-area network context. There are several other essential aspects to be considered. Scalability is definitely a major concern. In this paper, we have shown that our architecture significantly reduces consumed network bandwidth resources, but we have not discussed multicast state overhead. In general, if there are '*x*' MNs, each communicating with '*y*' CNs on average, then routers in the network should create '*x.y*' (*S,G*) states. A state, however, is created only en-route from the corresponding CN to MN.

---

[28] Our initial design does not require the functionality of the home agent (HA) or foreign agent (FA). However, we feel it is sometimes desirable to have the HA for accounting and security reasons. Moreover, presence of the HA renders Mobile IP networks capable of implementing our architecture with minor modification; namely, the support of multicast address in the *binding update*.

[29] The join refresh period for PIM-SM is typically 1 minute.

In terms of other scalability and robustness aspects, we retain properties of the underlying multicast routing infrastructure[3][26].

5.6. Sending Packets from the Mobile

To be able to send packets to a multicast address (whether to a multicast group or to another mobile node with a multicast address) the source address in these packets needs to pass the reverse path forwarding (RPF) check in the multicast routers. For that, the MN needs to use a unicast address that belongs to the network at which the MN is currently located. Obtaining such an address is similar to getting a care-of-address (through DHCP, or otherwise) in Mobile IP[30].

## 6. Related Work

Several architectures have been proposed to provide IP mobility support. Early work on IP mobility has been presented in[1][2]. Other, more recent, work by the IETF was done on Mobile IP (MIP)[4]. In MIP a mobile node (MN) is assigned a permanent *home* address and a *home agent* (HA) in its home subnet. When the MN moves to another *foreign* subnet, it discovers a foreign agent (FA) on that subnet and acquires a temporary care-of-address (COA) through a solicitation/advertisement message exchange with the FA. The MN informs the HA of its COA through a *registration* process. From that point on, packets destined to the MN's home address are sent first to the home network, are picked up by the HA and then are *tunneled* to the MN through the FA. This is known as the *triangle routing* problem, which is the major drawback of the basic MIP.

A proposed mechanism, known as *route optimization*, attempts to avoid triangle routing. In[5] route optimization is achieved by sending *binding updates*, containing the current COA of the MN, from the HA to the correspondent node (CN). In MIPv6[6][7] binding updates are sent from the MN to the CN with every move. Although this alleviates the triangle routing problem in MIP, the communication overhead is still high during handoff rendering MIP unsuitable for *micro* mobility[31] and causing it to be inadequate for audio applications.

Caching techniques are proposed in[8] and[6][32] to reduce packet loss during handoff. These techniques, in general, cause the old FA (before the move) to forward cached packets to the new FA (after the move) to recover packets that would otherwise be lost during the transition. The old FA needs to know the new COA of the MN before forwarding the cached packets; hence this technique still incurs handoff latency and results in out-of-order packets.

A hierarchical mobility management scheme was proposed in[13][14] that defines three hierarchical levels of mobility; local, administrative domain and global mobility. This scheme proposes to use MIP for the global mobility, while using *subnet foreign agents* and *domain foreign agents* for the other levels. It is not clear, however, how this hierarchy will be formed or how it adapts to network dynamics, partitions or router failures. The above study describes, implements and evaluates the local handoff protocol on the same subnet.

Another architecture for smooth handoff in ATM networks is proposed in[17]. This architecture uses a *virtual connection tree* (VC tree) that connects several radio cells within a region. The root of the tree for a given region handles virtual connections from and to the mobile nodes within that region. This root is statically allocated. It is unclear how this root is chosen and how the protocol reacts to network dynamics.

The Daedalus project[11][12] proposes to tunnel the packets from the HA using a pre-arranged multicast group address. The base station, to which the MN is currently connected, and its neighboring base stations (BSs), join that group and get the data packets over the multicast tree. Using beacons and signal strength measurements, the MN determines which BS should join the group and to which BSs it is likely to move in the near future. Advance buffering is used to achieve very low latency and reduced data loss. Experiments in a limited topology show that very low latency (< 15ms) and no data loss can be achieved up to 3-hop distance between BSs[33]. It is not clear how the scheme performs in larger wide-area topologies. This approach suffers from the triangle routing problem; packets are sent to the HA first and then to the MN.

---

[30] Using a unicast address may also solve some TCP interoperability problems, where current TCP implementations do not handle acks with class-D multicast address in the source field.

[31] Here, we use the term *micro* mobility (as opposed to *macro* mobility) to indicate small frequent movement, such as movement between neighboring offices or buildings. In the literature, it usually means movement within the same subnet or same domain.

[32] In MIPv6 this is called *router-assisted smooth handoff*.

An approach for providing mobility support using multicast (MSM-IP) is presented in[9][10]. In this approach, each MN is assigned a unique multicast address. Packets sent to the MN are destined to that multicast address and flow down the multicast distribution tree to the MN. This is similar, in concept, to the Daedalus project approach. However, it is not the HA that tunnels the packets using the multicast address, rather, it is the CN that sends packets directly to the multicast address. This approach avoids triangle routing, in addition to reducing handoff latency and packet loss by potentially using advance buffering. Experiments for MSM-IP were performed in a limited testbed consisting of three switched Ethernet subnets, and low packet loss and duplication were reported during handoff[34]. More work is needed to measure the protocol performance in larger networks.

In this paper, we present an architecture for supporting IP mobility. We avoid triangle routing and caching drawbacks suggested by the MIP architectures. Also, we avoid creating our own hierarchy of agents. Rather, we re-use the existing hierarchy and infrastructure of wide-area and inter-domain multicast routing[3][26]. We leverage off of some ideas offered by the Daedalus project and the MSM-IP to build our architecture and evaluate it in a wide-area networking context.

## 7. Conclusion and Future Work

This paper, to our knowledge, presents the most extensive wide-area evaluation of an IP mobility architecture. Our architecture is multicast-based, in which a mobile node is assigned a multicast address, and the correspondent nodes send packets to that multicast group. As the mobile node moves to a new location, it joins the multicast group through the new location and prunes through the old location. Dynamics of the multicast tree provide for smooth handoff, efficient routing, and conservation of network bandwidth.

We compare our architecture to the basic Mobile IP approach. Our simulation results show that Mobile IP consumes, on average, almost twice as much network bandwidth, experiences more than double the end-to-end delays and handoff latency, than does our architecture. Our measure of route efficiency *'r'*, and handoff latency *'L'* were found to be relatively insensitive to topology size (for the same topology type). *'r'* was also found to be relatively insensitive to movement patterns, while *'L'* was found to change according to the movement patterns. Our simulation scenarios used three movement models; *random, neighbor* and *cluster* movements, and 21 topologies of various sizes and degrees, synthesized by various commonly used methods of probabilistic generation and actual measurements.

We have presented a list of design issues, including start-up phase, host membership protocol and advanced joins. These issues need to be discussed in detail in our future work. We also plan to compare our architecture to other IP mobility support architectures, such as MIPv4 and v6 with route optimization and caching. In addition, we plan to conduct more detailed simulations to measure packet loss, jitter, delay and throughput during handoff. Another rich area of investigation includes studying the interaction with higher layer applications (such as TCP and multicast-based applications).

---

[33] Distance between BSs is defined as number of hops that multicast packets traverse between BSs.
[34] The authors point out some implementation problems, such as interaction between the class D multicast addresses and TCP, DHCP, ICMP and ARP. They propose suggestions to overcome some of these problems. The MSM-IP approach is similar, in concept, to our proposed approach in this study.